# Fine Tuning of Optical Transition Energy of Twisted Bilayer Graphene via Interlayer Distance Modulation

*Elena del Corro[1,\*], Miriam Peña-Alvarez[1,2], Kentaro Sato[3], Angel Morales-Garcia[4], Milan Bousa[1,5], Michal Mračko[6], Radek Kolman[6],Barbara Pacakova[7], Ladislav Kavan[1], Martin Kalbac[1], Otakar Frank[1,\*]*

[1] J.Heyrovsky of Institute of Physical Chemistry of the CAS, v.v.i., Dolejskova 2155/3, 182 23 Prague 8, Czech Republic

[2] Departamento de Química Física I, Facultad de Ciencias Químicas, Universidad Complutense de Madrid, 28040, Spain

[3] National Institute of Technology, Sendai College, Sendai 989-3128, Japan

[4] Department of Physical and Macromolecular Chemistry, Faculty of Science, Charles University in Prague, Hlavova 2030, Prague 2, 128 43, Czech Republic

[5] Department of Inorganic Chemistry, Faculty of Science, Charles University in Prague, Hlavova 2030, Prague 2, 128 43, Czech Republic

[6] Institute of Thermomechanics of the CAS, v.v.i., Dolejskova 1402/5, CZ 182 23 Prague 8, Czech Republic

[7] Institute of Physics of the CAS, v.v.i, Na Slovance 2, 182 21, Prague 8, Czech Republic

Twisted bilayer graphene (tBLG) represents a family of unique materials with optoelectronic properties tuned by the rotation angle between the two layers. The presented work shows an additional way of tweaking the electronic structure of tBLG: by modifying the interlayer distance, for example by a small uniaxial out-of-plane compression. We have focused on the optical transition energy, which shows a clear dependence on the interlayer distance, both experimentally and theoretically.

# I. INTRODUCTION

Since the advent of graphene, tuning of its electronic structure has been one of the strongest focal points for many researchers. However, so far the vision of exploiting the unique properties of graphene for the replacement of silicon in electronics has been hampered by the inability to open a sizeable band gap in a simple, controlled and cost-effective manner [1]. For this purpose, bilayer graphene (BLG) holds more promise, for applications such as nanoelectronics, than monolayer graphene, as it offers several routes of profiting from the interactions between the two layers [2-4], e.g. by dual gating [4,5], molecular doping [6] or theoretically by mechanical deformation [7]. Similarly, the appealing concept of a Bilayer Pseudo-Spin Field Effect Transistor (BiSFET) still exists only at the theoretical level [3,8,9]. The interlayer distance could be one of the important parameters controlling the excitonic gap in BiSFET [10].

Twisted bilayer graphene (tBLG), i.e. a system where the alignment of the two graphene layers deviates from the periodic (AB) Bernal stacking, has recently attracted increased attention. In fact, each tBLG with a particular twist angle represents a unique material in terms of its optoelectronic properties [11]. The relative rotation of the layers leads to the formation of a superlattice, which have manifested themselves as Moiré patterns in high-resolution transmission electron microscopy [12] or scanning tunneling microscopy [13] studies. Importantly, the interference in superlattices gives rise to van Hove singularities (vHS) in the density of states (DOS), with their energy gap dependent on the twist angle [12-14]. The vHS cause optical coloration of the tBLG [11] and a strong resonance enhancement of the Raman G mode when the laser excitation matches the vHS energy [12,15,16]. As mentioned above, the interaction between the two layers depends on the twist angle; however, the influence of the interlayer distance has never been examined. Moreover, the interlayer space in stacked two dimensional materials provides an additional feature through the so-called van der Waals pressure acting upon molecules or crystals trapped in between the layers [17,18]. However, this phenomenon is still to be fully explained.

In the presented work, we have studied tBLG of various origin under direct uniaxial out-of-plane compression in a low stress regime. Simultaneous in-situ Raman spectroscopy measurement revealed a clear modulation of G band enhancement, indicating changes in the resonance conditions and hence in the energy of the vHS. In order to evaluate the effect of compression, we have performed theoretical calculations of the DOS of tBLG, modulating both the interlayer and the in-plane C-C distance. Our calculations reveal variations of the vHS energy as a dependence of the interlayer distance, as large as 200 meV, while negligible variations are detected when decreasing the $a$ lattice parameter.

## II. MATERIALS AND METHODS

*Experiments.* Single layer graphene samples of $^{12}C$ and $^{13}C$ were prepared using the CVD method, described elsewhere [19]. Labeled bilayer graphene was obtained by sequential transfer of individual monolayers from copper foil onto a sapphire disc, using the reported wet transfer method with polymethylmethacrylate [20]. Additionally, as-grown and exfoliated $^{12}C$ BLG samples were studied for comparison. The experimental setup consisted of a gem anvil cell coupled to a Raman spectrometer (LabRAM HR, Horiba Jobin-Yvon). In order to perform direct out-of-plane compression, a modified sapphire cell was used, with one anvil substituted by a sapphire disc containing the sample. In such conditions, the use of a conventional stress marker is inadequate and therefore, stress was estimated from the evolution of the Raman peaks of sapphire [21]. Raman spectra and maps were registered using an $Ar^+/Kr^+$ laser working at 488.0 nm, 514.5 nm, 532.0 nm and 647.1 nm, keeping the power on the sample below 1 mW. A 50x objective produced a laser spot on the sample of ~1 μm diameter. Grating with 600 grooves $mm^{-1}$ was used to provide spectral point-to-point (pixel) resolution of ~1.8 $cm^{-1}$ at 488.0 nm excitation wavelength. After each 0.1 and 0.5 GPa compression step, single spectra and Raman maps (20x20 $μm^2$, 1-2 μm sampling steps) were registered, respectively, for selected sample grains fulfilling the resonant conditions at the corresponding laser excitation energy. All peaks the spectra are fitted with Lorentzian lineshapes.

*Calculations.* A commensurate structure of tBLG is characterized by two integers (n,m), which define the rotation angle between the layers. In the calculation, we use primitive vectors $\boldsymbol{T}_1$=n$\boldsymbol{a}_1$+m$\boldsymbol{a}_2$ and $\boldsymbol{T}_2$=(n+m)$\boldsymbol{a}_1$-n$\boldsymbol{a}_2$ for (n,m) tBLG [14]. Here $\boldsymbol{a}_1$=$a$(√3/2,1/2), $\boldsymbol{a}_2$=$a$(√3/2,-1/2), $a$=|$\boldsymbol{a}_1$|=|$\boldsymbol{a}_2$| are the primitive vectors and lattice constants for monolayer graphene, respectively. The electronic structure and DOS of tBLG are calculated using the tight binding method [22,23] with a different interlayer distance and in-plane lattice constant in order to evaluate the effect of compression. The adopted tight binding parameter is a function of the distance between carbon atoms [22]. The optical transition energy, corresponding to the enhancement effect of Raman intensity, is determined from the results. Optical transition between the saddle points of electronic structures is not allowed [24].

## III. RESULTS AND DISCUSSION

The Raman spectrum of labeled BLG has been previously reported [25]. The phonon frequency ($\omega$) scales inversely with the atomic mass and, therefore, the Raman peaks from each layer can be distinguished [26]. Thus, the Raman spectrum of labeled BLG is dominated by four peaks: two G bands (1525 and 1590 cm$^{-1}$) and two 2D bands (2620 and 2710 cm$^{-1}$), with the lower frequency peaks corresponding to $^{13}$C [26]. Analogously, in the case where lattice disorder is present in the sample, two D bands appear at ~ 1303 and 1347 cm$^{-1}$ (2.54 eV excitation energy).

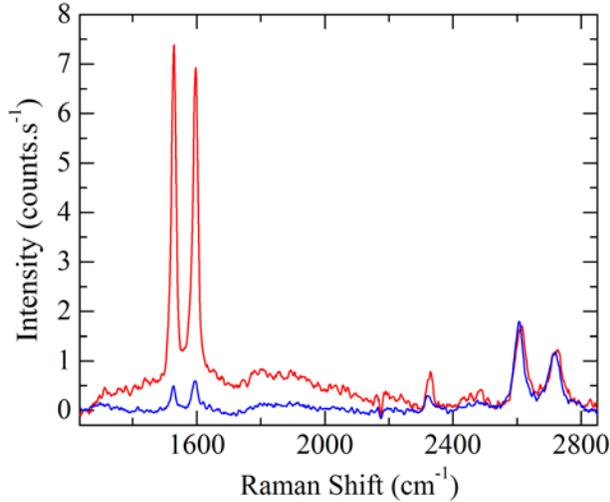

FIG. 1. Raman spectra of the $^{13}$C/$^{12}$C tBLG at ambient conditions, with and without G band enhancement at the laser excitation wavelength of 488.0 nm. The spectra are normalized to the 2D band amplitude.

As stated above, tBLG shows vHS in the density of states, with their energy gap dependent on the twist angle. When the excitation wavelength matches the energy difference between these vHS (the optical transition energy), an enhancement of the G band intensity is observed, and the G/2D intensity ratio (amplitudes) increases by a factor of 15, see Figure 1, or even more depending on the sample. In order to locate tBLG grains in resonance with the excitation energy used, Raman maps of 40×40 μm² were measured with distinguishable regions exhibiting enhanced G band (see Figure S2). In Figure 2a, we show single Raman spectra acquired at the center of a tBLG grain, fulfilling the resonant conditions at 2.54 eV excitation energy, i.e. ~13 ° rotation angle [12], with sequentially increasing out-of-plane compression up to approximately 1.6 GPa. The behavior under compression over a larger stress range is presented and discussed in the Supporting Information [27]. The use of isotopically labeled tBLG allows us to bring to light any potential disharmony in the evolution of the two layers during the high stress experiment. However, Figure 2a shows that both layers in the examined tBLG (and in all other experimental runs mentioned further) manifest the same behavior.

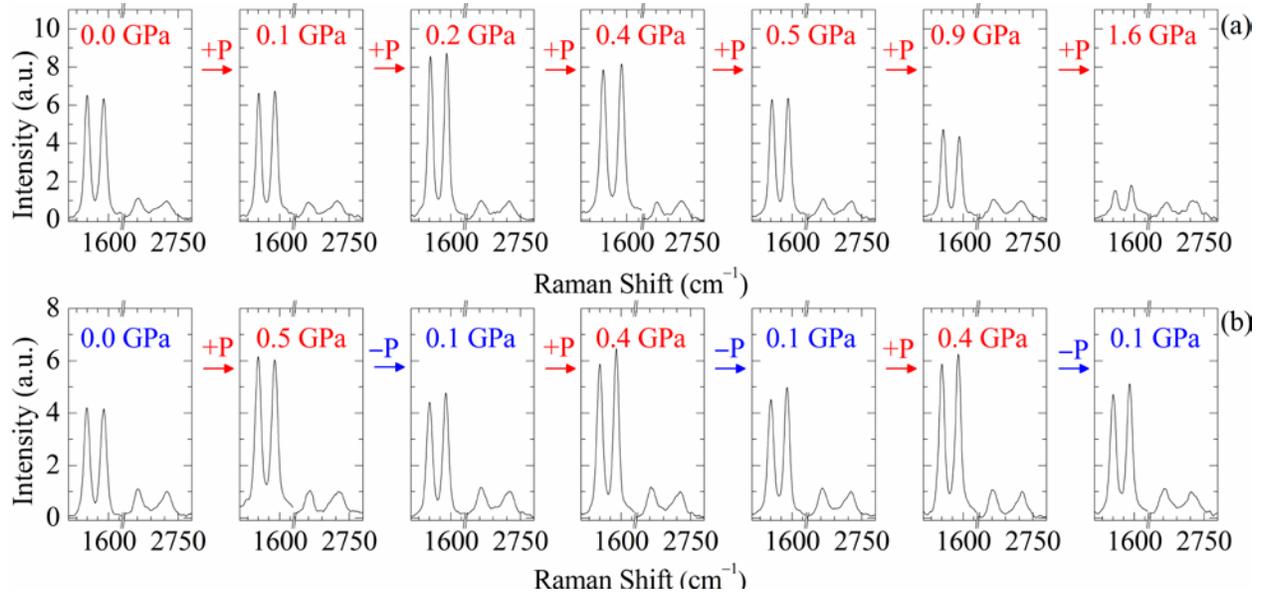

FIG. 2. Evolution of Raman spectra of $^{13}C/^{12}C$ tBLG (a) with compression up to 1.6 GPa, and (b) with stress cycling of approximately ±0.4 GPa. The laser excitation wavelength is 488.0 nm. The spectra are normalized to the 2D band amplitude in each plot.

As shown for graphite [28] and other layered materials [29], out-of-plane compression decreases, preferentially, the distance between layers. In other words, the out-of-plane compressibility is much higher than the in-plane compressibility, owing to weak interlayer forces. At the initial stages of compression (up to 0.4 GPa), we observe an increase in the G band enhancement by a factor of *ca.* 1.5 (the increase in the G band enhancement can reach a factor of 4, depending on the sample, see Figure 3a), which is reduced to the original value at about 1 GPa. Such a variation indicates that the decrease in the interlayer distance modifies the resonance conditions, *i.e.* the electronic properties of the system. During the later stages of compression above 1 GPa, the enhancement continues to decrease due to other factors affecting the compression experiment, such as the sample disorder, as reported before [30] and discussed in the Supporting Information, Figure S3 [27].

The G band enhancement within the 0.5 GPa range warrants further investigation. In Figure 2b, we present the stress performance during consecutive compression cycles of a tBLG grain. We have observed that the behavior described above is reversible in the low stress range. Such reversibility confirms the assumption that the change in the interlayer distance is the main factor of the modulation of the tBLG electronic properties. While the charge doping has been shown to modify the electron resonance in tBLG [31], a possible change in the doping state induced by an irreversible purging of impurities (remnant from the transfer process) from the interlayer space, can be ruled out based on the reversible behavior shown in Figure 2b. A change of rotation angle with shear stress can also be excluded as an explanation of the G band enhancement in view of Figure 2b. Moreover, additional observations based on the Finite Elements (FE) method, compiled in the Supporting Information [27], demonstrate only insignificant shear stresses, at least three orders of magnitude smaller than the out-of-plane compressive forces, hence no appreciable relative movement of the layers is expected.[32]

In order to prove the universality of our findings, additional experiments were performed by employing several excitation energies and analyzing different tBLG samples. In Figure 3a, we present the evolution of the G/2D amplitude ratio with increasing stress for labeled tBLG with different twist angles; each of them excited with the corresponding resonant laser energy in order to observe G band enhancement (for the same plot, but with ratio of G/2D integrated areas, see Figure S11 in the Supporting Information [27]). We observed that the variation of the resonant conditions is qualitatively the same regardless of rotation angle and excitation energy. Grains with different twist angles exhibit different magnitudes of energy band-gap modulation, as evidenced by the varying enhancement factor and position of the enhancement maximum for individual excitation wavelengths in Figure 3a, and as demonstrated by our theoretical calculations in the last section.

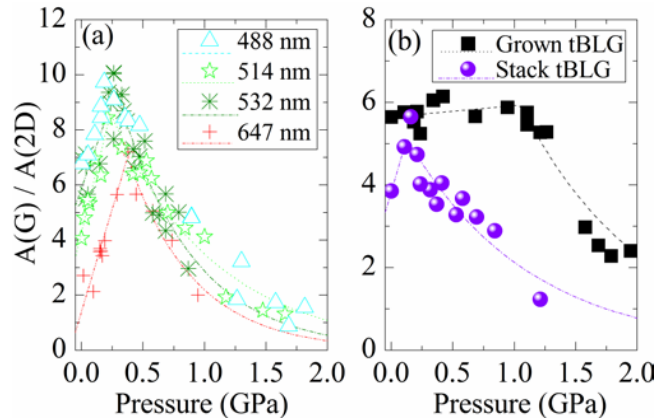

FIG. 3. Evolution of the G/2D amplitude ratio with increasing stress: (a) Labeled tBLG with rotation angles of 13.0, 12.3, 11.9 and 9.8° excited with 488.0, 514.5, 532.0 and 647.1 nm laser wavelengths, respectively, to achieve G band enhancement in each sample [12]. (b) $^{12}$C CVD

BLG (squares) and $^{12}$C/$^{12}$C stack BLG (circles); both measured in a region resonant with 488.0 nm excitation wavelength. Dotted lines are guides for the eye.

In Figure 3b, we compare the compression behavior of as-grown $^{12}$C tBLG with that prepared by a sequential transfer of two monolayers. The latter shows a compression behavior analogous to the labeled tBLG. Note that before compression, the sample consists of two layers of graphene sequentially transferred, in which remnant polymers or other impurities from the transfer process may hinder a closer contact between them (see AFM analysis in the Supporting Information, Figure S10 [27]). However, in the as-grown tBLG samples, the graphene layers are in closer contact in their original state, and therefore the increase in G band enhancement is almost absent. Note that the maximum stress-induced enhancement in the sequentially transferred tBLG is the same as the initial enhancement in the as-grown tBLG.

To further prove the effect of interlayer distance on the resonant conditions and therefore on the electronic properties of tBLG, we performed a single compression cycle up to 0.6 GPa on a tBLG grain with a G band enhancement manifested for both the 488.0 and 514.5 nm excitation wavelengths. Note that the G band enhancement is not the maximum possible for either of the excitation energies, which indicates that the rotation angle of the measured grain is between 13.0 and 12.3 degrees (see Figure 4). When increasing the stress, thereby reducing the interlayer distance, resonant conditions are changed in such a way that the same grain more closely fulfils resonance conditions for 488.0 nm, while simultaneously, enhancement is almost lost for the 514.5 nm line. This result indicates that the optical transition energy for the analyzed tBLG grain has increased upon compression, abruptly moving closer to 2.54 eV (488.0 nm) rather than to 2.41 eV (514.5 nm), followed by a stabilization of the energy after applying a stress of ~0.5 GPa.

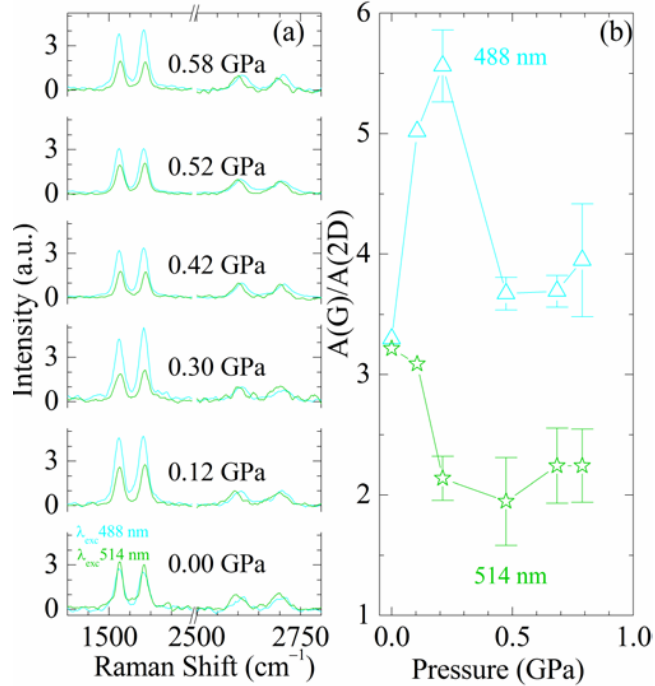

FIG. 4. Evolution of Raman spectra of $^{13}C/^{12}C$ tBLG with compression up to 0.6 GPa. (a) The same tBLG grain is measured with 488.0 (blue spectra) nm and 514.5 nm excitation wavelengths (green spectra). Each spectrum is normalized to the 2D band amplitude. (b) G/2D amplitude ratios from (a) as a function of increasing stress.

Finally, modulation of the electronic structure of tBLG in the low stress regime, demonstrated by G band enhancement, is examined with the aid of calculations based on the tight-binding method. As an example, we present the DOS of the (3,2) tBLG structure in Figure 5a (corresponding to a ~13° rotation angle, in resonance with 488.0 nm excitation wavelength [14]) at selected interlayer distances ($d$) ranging from 0.50 to 0.25 nm. We scan a wide $d$ interval in order to account for all possible experimental variables: the decrease of $d$ by means of stress and the increase of $d$ in pristine samples due to their preparation method [27]. We observed that both the DOS and the vHS energy are modified by $d$. Specifically, in Figure 5b, we see that upon decreasing the interlayer distance to ~0.45 nm, the optical transition energy quickly reaches a maximum, and then starts to decrease at a slower pace as layers continue to move closer to each other. This is in perfect agreement with the experimental results presented in Figure 4, where the initially similar resonance (with 2.41 and 2.54 eV) laser excitations is immediately moved towards 2.54 eV and then steadied. Along the whole analyzed range of interlayer distances, the optical transition energy shows variations as large as 200 meV, for the given twist angle. This variation of optical transition energy with $d$ alters the resonance conditions, thereby explaining the experimental observations where the G band enhancement changes with stress at that particular excitation energy. The theoretical behavior of (1,5) tBLG, which is slightly further from the resonance with a 488.0 nm laser (twist angle of ~15°), has also been checked; the optical transition energy follows a similar trend as in (3,2) tBLG, albeit with a smaller energy difference up to 80 meV.

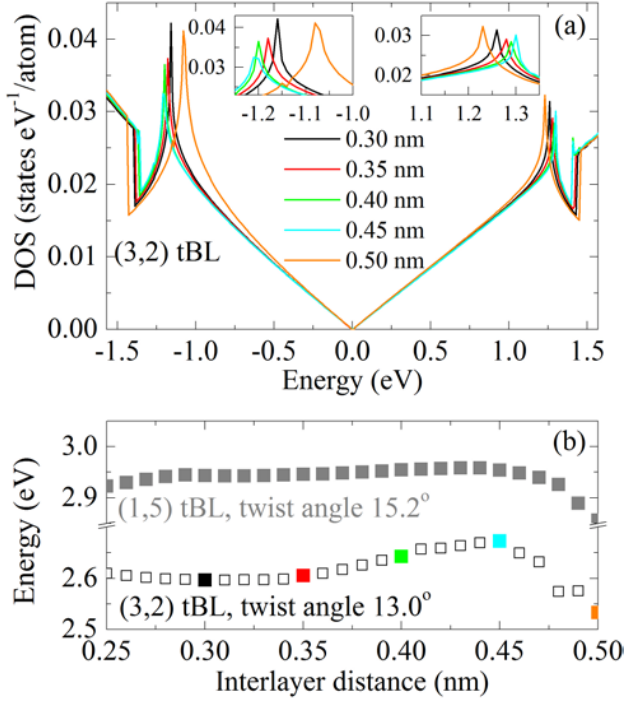

FIG. 5. (a) Density of states of (3,2) tBLG at different interlayer distances. (b) Optical transition energy as a function of the interlayer distance for (3,2) and (1,5) tBLG, labeled with squares and circles, respectively. Highlighted points in the (3,2) tBLG curve in (b) correspond to the curves in (a) with the same color.

As noted above, despite being lower, the tBLG out-of-plane compression demonstrates clear in-plane compressibility [33], proved by the blue-shift of the Raman spectrum [27]. The influence of the in-plane compression on the electronic properties of tBLG is presented in Figure 6. DOS was calculated for the same (3,2) tBLG structure at a fixed interlayer distance (0.35 nm) with decreasing $a$ lattice parameter values, from the equilibrium, 0.2460 nm, down to 0.24565 nm (corresponding to an in-plane compression of >1.5 GPa, obtained from the equation of state of graphite [34]). Within this $a$ range, only a negligible increase of vHS, ~1meV, is observed.

Moreover, when decreasing the *a* parameter more drastically to 0.2435 nm (estimated stress of ~10 GPa), the increase in vHS energy obtained theoretically is lower than 10 meV, which is 20 times smaller than is observed for interlayer distance modulation. Thus, according to our calculations, in-plane stress can be discarded as the origin of the modulation of the resonance conditions in tBLG.

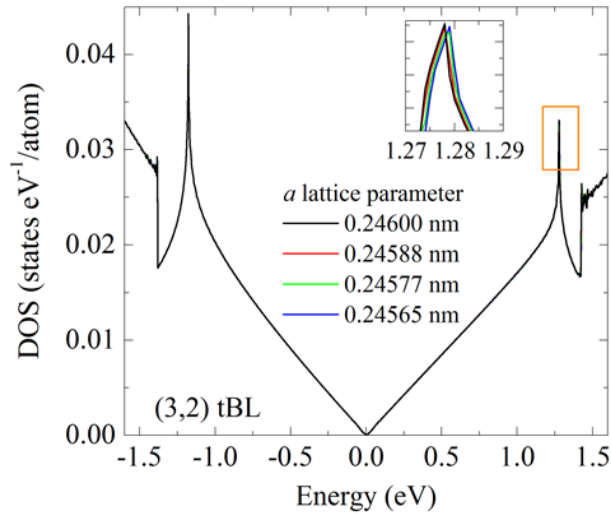

FIG. 6. The density of states of the (3,2) tBLG, with 0.35 nm interlayer distance, as a function of the *a* lattice parameter.

## IV. CONCLUSIONS

In summary, we have shown the dependence of the energy of van Hove singularities with interlayer distance in tBLG. Various samples of tBLG were subjected to out-of-plane compression and their behavior was monitored in-situ by Raman spectroscopy using different laser excitation energies. The experiment showed a change in Raman G band enhancement reflecting the modified resonance conditions caused by the altered vHS energy. The results were corroborated by tight binding calculations, which revealed an initial increase in optical transition energy upon decreasing the interlayer distance to ~0.45 nm, where the maximum energy was reached, followed by a gradual decrease with further narrowing of the interlayer gap. The calculations also showed that the in-plane compression of graphene layers was not responsible for the changes in optical transition energy. The sensitivity of tBLG to interlayer distance can prove valuable in optoelectronic applications, and based on our observations, it also explains the differences in the magnitude of the G band enhancement observed at particular laser excitations for differently prepared samples.

## ACKNOWLEDGMENTS

This work was funded by the Czech Science Foundation (No. 14-15357S), MSMT ERC-CZ project (LL 1301) and European Union H2020 Program (No. 696656 – GrapheneCore1). K.S. acknowledges support from MEXT (Grant 23710118). M.M and R.K. acknowledge the Czech Science Foundation (GACR 16-03823S) within the institutional support RVO 61388998.

# Fine Tuning of Optical Transition Energy of Twisted Bilayer Graphene via Interlayer Distance Modulation

*Elena del Corro, Miriam Peña-Alvarez, Kentaro Sato, Angel Morales-Garcia, Milan Bousa, Michal Mračko, Radek Kolman, Barbara Pacakova, Ladislav Kavan, Martin Kalbac, Otakar Frank*

**High Pressure behavior of BLG**

In general, compressive stress leads to the stiffening of phonons, which is translated into an upshift of the Raman spectrum [S1]. In Figure S1, we present $\omega_{2D}$ as a function of $\omega_G$ for the $^{12}C$ components, comprising data coming from the three different sample types. We observe that for all the samples, the frequency correlation 2D/G follows a linear evolution with a slope of ~2.2, indicating a purely biaxial strain effect, in absence of doping phenomena [S2]. As expected, no differences are observed between samples with $^{12}C$ as the top or the bottom layer - once the cell is closed, both layers are equally in contact with the sapphire surface.

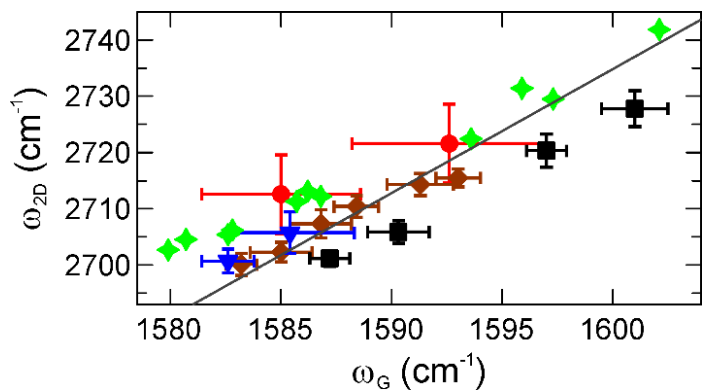

**Figure S1.** Frequency correlation 2D/G. Each data point represents the average frequency from a Raman map (*ca.* 100 spectra, error bars for the standard deviation). Different samples: exfoliated BLG (green stars), tBLG with $^{12}C$ top (red circles) or bottom layer (black squares, blue triangles and orange diamonds).

In Figure S2, we present the evolution of the Raman map of a tBLG around a graphene grain with stress, in resonance with the 2.54 eV laser excitation energy. In the low stress regime, we observe an analogous modulation of the enhancement as that discussed in the main manuscript. With the stress increasing over 2 GPa, the enhancement of the G band is completely suppressed and the whole Raman map shows similarly low G/2D intensity ratio in all areas. The loss of enhancement is completely irreversible and it is accompanied by a sudden pronounced increase of the D mode bands. A more detailed analysis is shown in Figure S3 where the D/2D intensity ratio is presented with increasing stress (each point represents the average value of a Raman map). We observe that the amount of disorder, directly related to the D band intensity, remains constant at the first stages of compression, but sharply increases when the stress exceeds the 1.5 GPa threshold, exactly coinciding with the disappearance of the G band enhancement. A similar bleaching of the G band enhancement associated with an increased disorder has been recently observed in an ion-irradiated tBLG [S3].

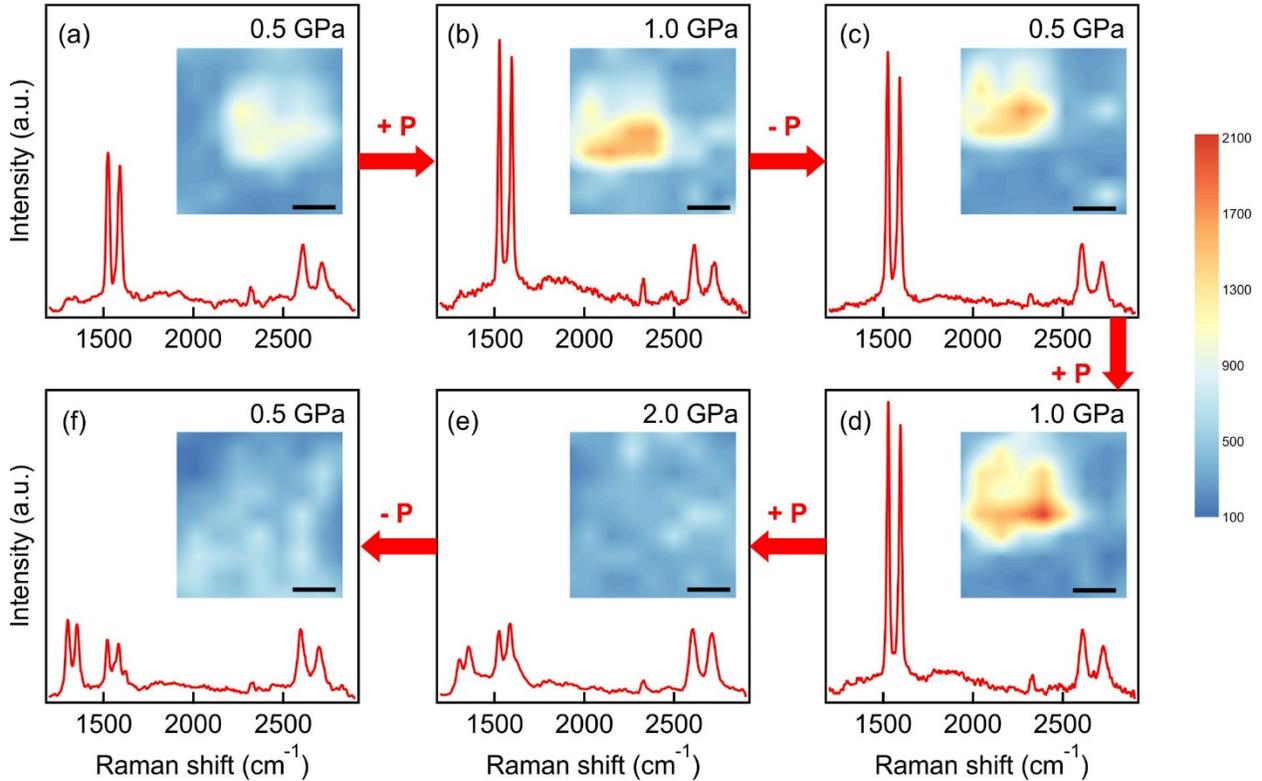

**Figure S2**. The evolution of the Raman spectra of $^{13}C/^{12}C$ tBLG with stress in approximately the same region of the sample (small spatial fluctuations are caused by the lower visibility of the sample through the anvil). The maps in the insets show the absolute G band intensity, and the corresponding spectrum selected from the region with the highest G band enhancement. The scale bar is 10 μm.

Figure S3 presents the intensity of the D band over the intensity of the G band (both related to the 2D band). Data points from each particular Raman map (at a given pressure) always follow a linear trend, showing an uniform defect distribution across the map, with only negligible changes in the pressure range up to 1.5 GPa. The green points in Fig. S3b, corresponding to the 1.5 GPa pressure, also show the increase of the I(G)/I(2D) ratio, i.e. the G band enhancement. The increase of pressure from 1.5 to 2.0 GPa is accompanied by a sudden increase in the slope of the curve (hence the I(D)/I(G) ratio).

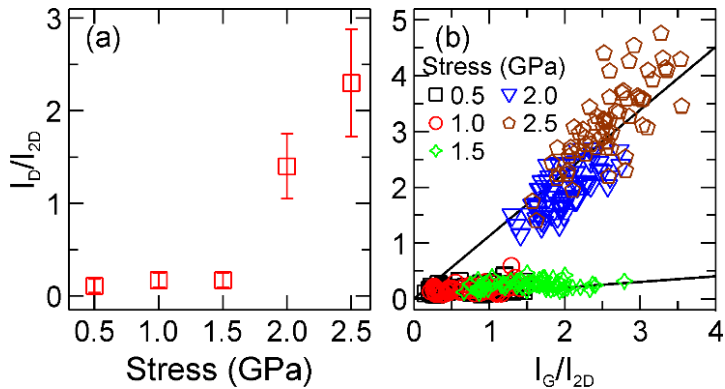

**Figure S3**. (a) D/2D intensity ratio evolution with stress. Points represent the average values from Raman maps (error bars for the standard deviation. (b) D/G Intensity correlation. Solid lines for the linear fitting function of merged data, 0.5-1.0-1.5 and 2.0-2.5 GPa, with slopes of 0.10 and 1.13, respectively.

## Compression anvil experiments

The compression anvil experiments have been modelled using Finite Elements (FE) method and compared to Raman mapping of the whole contact area between the sapphire anvil, with 350 μm cullet, pressing onto a 1 cm (0001) sapphire disc. The Raman band of a sapphire at ~417 cm$^{-1}$ has been used to determine the out-of-plane contact stress [S4]. The map of thus measured stress is shown in Figure S4. WITec Raman spectrometer coupled to a confocal microscope with a 100x objective, using a 1800 lines/mm grating and laser excitation at 532 nm, was used for the mapping.

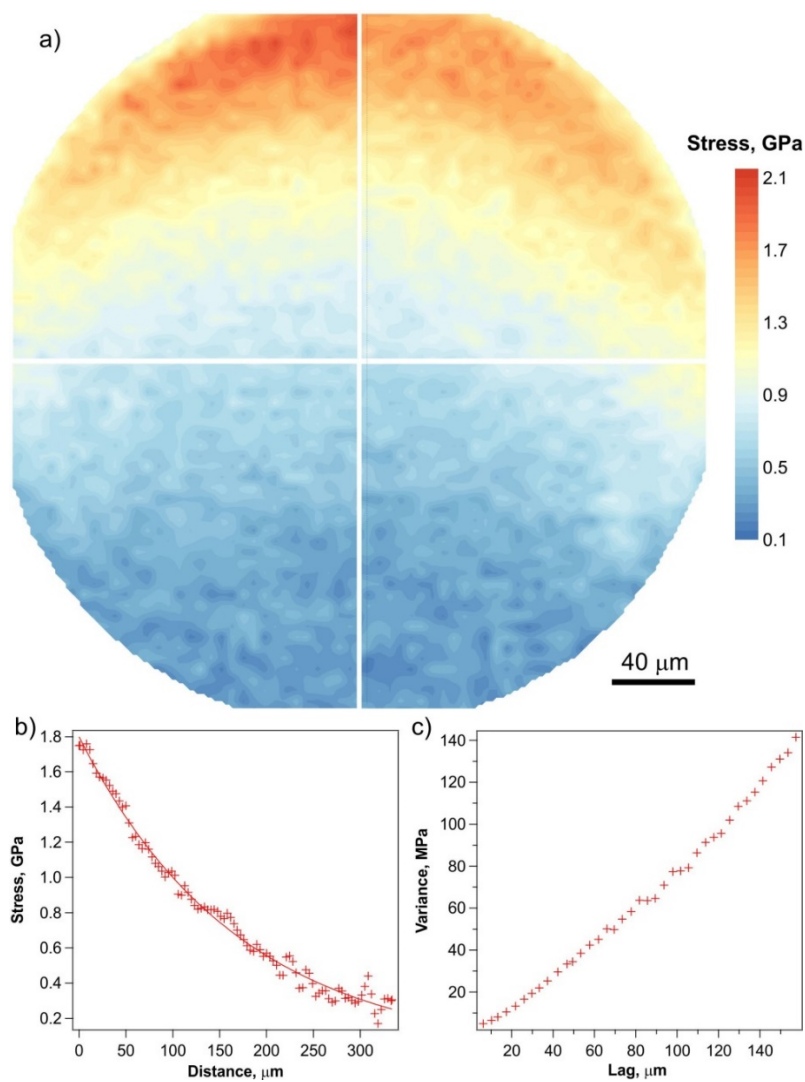

**Figure S4**. Uniaxial out-of-plane stress distribution in the contact plane between sapphire anvil and disc (a), stress profile from the top to the bottom of the map (b), and variance of stress calculated for the whole map (c).

The map (Fig. S4a) and the profile extracted from it (Fig. S4b) clearly show that the anvil and the disc are not in a perfectly parallel alignment, but at a certain tilt, creating an exponential stress gradient from one edge of the cullet to the opposite one. A very similar behavior of the stress profile can be observed in the FE simulations next, however, with different stress magnitudes in the maxima on the edge for the disc and the anvil. We can assume that the Raman signal in Fig. S4 comes from both the disc and the anvil, which in turn would lead to a substantial peak broadening. However, the FWHM of the sapphire 417 cm$^{-1}$ band remains in the range 3-5 cm$^{-1}$, only with an abrupt increase towards the edge with highest stress (Figure S5).

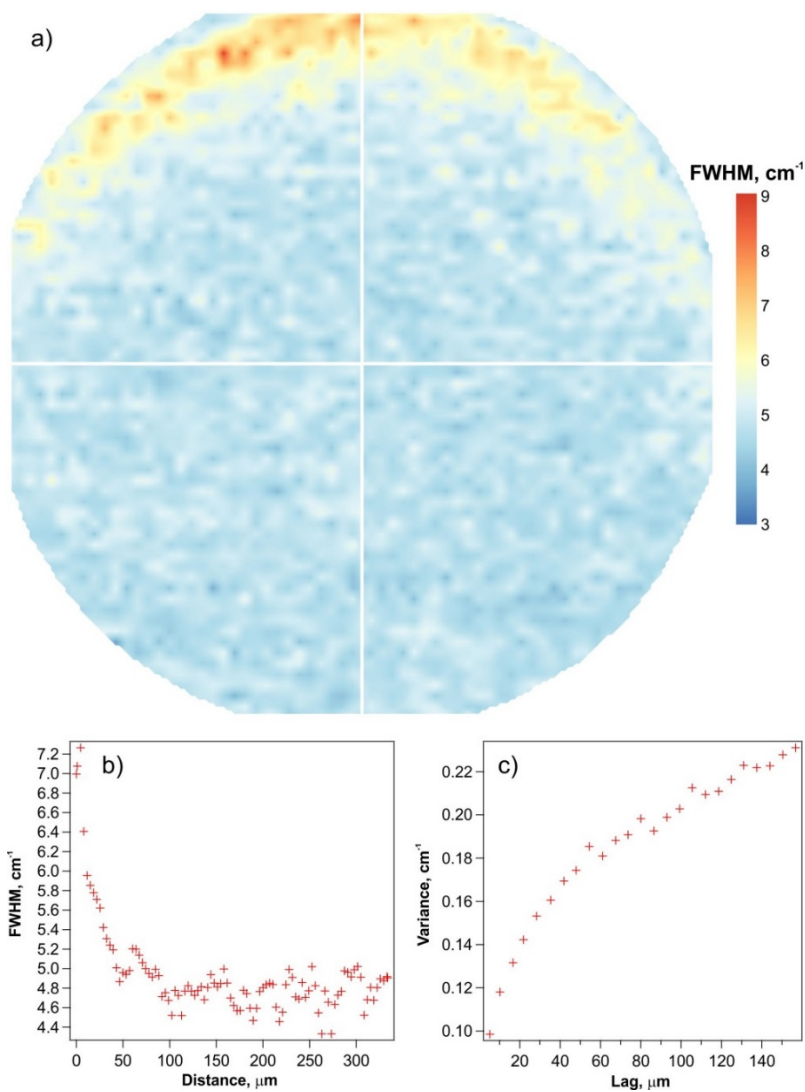

**Figure S5**. FWHM of the 417 cm$^{-1}$ band in sapphire at the contact area between the anvil and the disc (a), FWHM profile from the top to the bottom of the map (b), and FWHM variance calculated for the whole map (c).

On the other hand, the middle of the cullet, where all the actual measurements presented in the main text took place, shows smaller changes, see Figure S6. The highest change in stress is observed again along the top-bottom direction, with a slope of ~4.2 MPa/µm. For Raman maps of the graphene specimens, regions of 40x40 µm$^2$ were measured, hence the largest change caused by the stress gradient would amount to 168 MPa from one edge of the mapped area to the other. Given the spectral point-to-point resolution of the spectrometer (0.75 cm$^{-1}$) and the shift rate of the sapphire band (2.1 cm$^{-1}$/GPa), it is obvious that the stress gradient across the graphene map is smaller than the resolution of the spectrometer. In graphene, such a gradient would be reflected by the Raman G bad shift of ~1.5 cm$^{-1}$. We also note that the variance (i.e., average variation of values between any two points at a particular Lag distance, Fig.

S6c) in the middle part of the anvil is only ~ 8 MPa at 40 μm Lag. The much smaller value than the gradient itself is caused by only negligible variations in the left-right direction in the stress map (Fig. S6a). Additionally, the evolution of the widths of the sapphire bands in the same area of the cullet center (Fig. S6d-e) shows no gradual change of the FWHM in the top-bottom direction, only a change of ± 0.2 cm⁻¹, again smaller than the spectral resolution of the Raman system. To elucidate possible stress distribution and directions in the anvil, we have conducted FE simulation using force and tilt resembling the gradient presented in Figures S6 and S7. The calculated distribution of the contact pressure is plotted in Figure S7 for the disc (a) and the anvil (b), along with the stress gradient depicted as profiles in Fig. S7c and e, for the disc and the anvil, respectively (with their average in (d)).

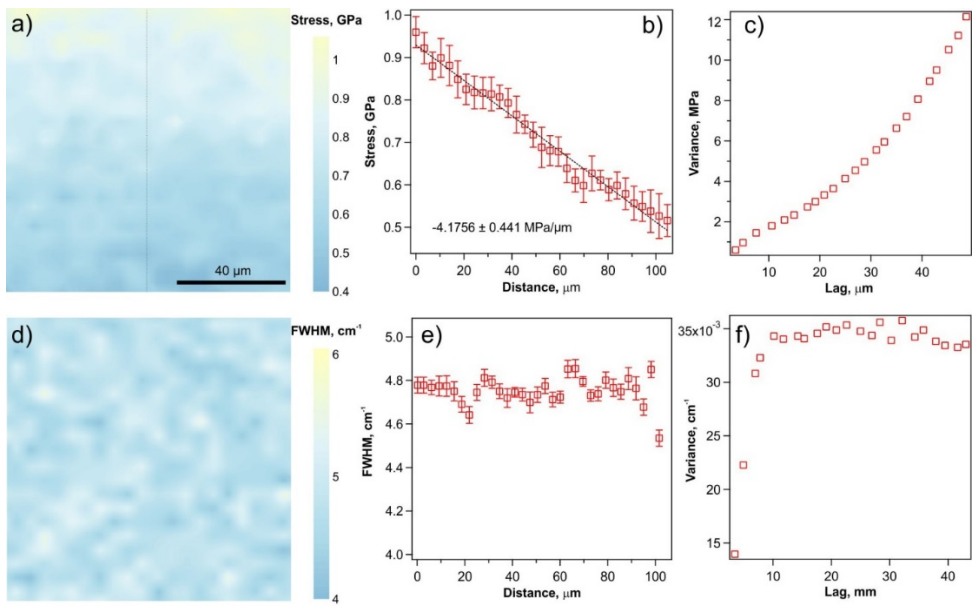

**Figure S6.** Uniaxial out-of-plane stress distribution in the center of the contact plane between sapphire anvil and disc (a), stress profile from the top to the bottom of the map (b), and variance calculated for the whole map (c).

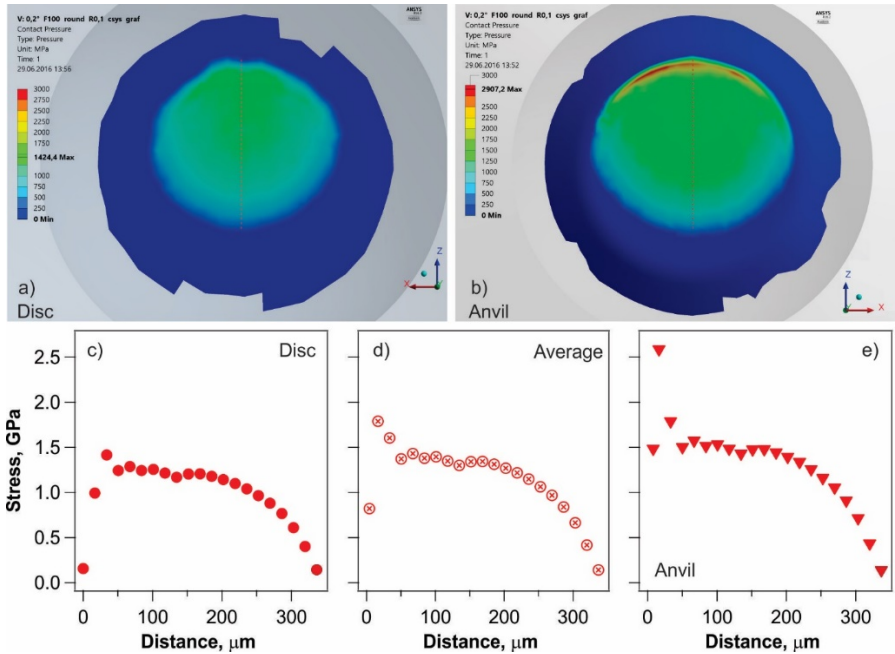

**Figure S7.** Contact pressure distribution in the sapphire disc (a) and anvil (b) at 0.2° tilt configuration, and the stress profiles taken from top to the bottom of the map for the disc (c) and anvil (e) with their average in (d).

The difference in the calculated contact pressure in the disc and the anvil is in the most of the contact area smaller than ~250 MPa, however, reaches up to 1.6 GPa at the edge. Such a discrepancy qualitatively corresponds to the observation of the sudden sapphire Raman peak broadening (Fig. S5). We also note that the evolution of the contact pressure along the profile follows a more complicated trend than in the experiment, which is probably caused by the setting of the anvil cullet as slightly rounded (see further for FE simulation details) to avoid singularities in the calculation. Furthermore, the contact pressure is higher than in the experiment. Nevertheless, the calculated stress gradient is not larger than the one measured.

We also extracted the principal stress directions from the FE simulation. The results from the middle of the contact plane for the 0.2° tilt, along with a less tilted (0.1°) and a fully parallel configuration (0°) are shown in Table S1. The Euler angles show the presence of shear components, however, there are random variations also for the parallel configuration. All the principal stress components are compressive in nature, with the minimum stress direction (i.e., maximum compression) always perpendicular to the contact plane, and the other two directions are in-plane, perpendicular and close to each other in magnitude. The ratio of the value at the minimum to the value at the maximum stress direction is approx. 0.55 ± 0.02, in all configurations, which is in a very good agreement with previous experimental results

[S5]. Additionally, there are slight differences between the stresses in the disc and the anvil, however, there is no trend going from parallel to 0.1° and 0.2° tilt, with the variations kept randomly within 1-7%.

**Table S1.** Principal stress directions in the anvil/disc at parallel, 0.1 and 0.2° tilt configuration. Note that for the Euler angles, the initial coordinate system has the XZ plane parallel with the contact plane. The ~30° $\theta_{yz}$ shows only the rotation around the vertical axis, without any physical meaning due to the rotational symmetry considerations.

| Angle | Body | Principal Stresses | | | | Euler Angle [°] | | |
|---|---|---|---|---|---|---|---|---|
| | | $\sigma_1$ | $\sigma_2$ | $\sigma_3$ | $\sigma_1/\sigma_3$ | $\theta_{xy}$ | $\theta_{yz}$ | $\theta_{zx}$ |
| | **Disc** | -694.0 MPa | -755.3 MPa | -1276.9 MPa | 0.54 | -90.8 | 32.5 | -90.0 |
| **0°** | **Anvil** | -739.4 MPa | -804.7 MPa | -1293.6 MPa | 0.57 | -87.6 | 34.9 | 89.8 |
| | Anvil/Disc | 1.07 | 1.07 | 1.01 | | | | |
| | **Disc** | -687.6 MPa | -752.0 MPa | -1291.7 MPa | 0.53 | -92.2 | 30.3 | 93.6 |
| **0.1°** | **Anvil** | -693.3 MPa | -760.0 MPa | -1253.8 MPa | 0.55 | -91.0 | 35.5 | 87.1 |
| | Anvil/Disc | 1.01 | 1.01 | 0.97 | | | | |
| | **Disc** | -666.4 MPa | -734.4 MPa | -1254.2 MPa | 0.53 | -91.6 | 27.4 | 94.4 |
| **0.2°** | **Anvil** | -692.8 MPa | -753.5 MPa | -1234.1 MPa | 0.56 | -86.9 | 28.1 | 84.5 |
| | Anvil/Disc | 1.04 | 1.03 | 0.98 | | | | |

Finally, the friction between the disc and the anvil has been calculated for the tilted contact (Figure S8), clearly showing a negligible relative movement between the two planes (note the scale in MPa, in contrast to the scale in GPa in Figures S3-S6).

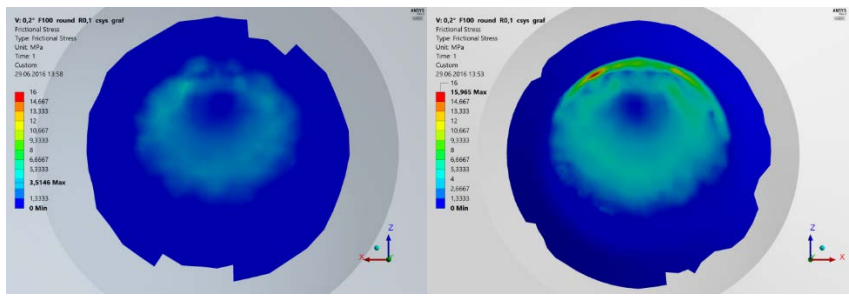

**Figure S8**. Map of friction stress (tangential stress in contact) between the disc (a) and the anvil (b) disc calculated for the 0.2° tilt configuration.

**Simulation method** In this Finite Elements analysis, a sapphire anvil was pressed against a sapphire disc, considering different anvil/disc orientations (parallel, 0.1° and 0.2° tilted). The geometry of the anvil is shown in Figure S9a; the sapphire disc is a rotational cylinder with diameter of 10 mm and height of 2 mm. Computation could not be performed on ideal geometry because of the singularity on contact surface edge [S6]. Therefore, the geometry of the cullet had to be slightly modified, also better representing the real situation. The flat face of the cullet was modelled as spherical with a radius of 20 mm and the edge of the cullet was also rounded, with a radius of 0.05 mm. Sapphire is an anisotropic material with trigonal symmetry. In our analysis, the c-axis of the crystal is oriented parallel to the rotational axis. The stiffness coefficient in the c-axis direction is represented by $c_{33}$. According to axis-symmetry, the rotation of other axes is not important. Unstructured mesh was used for analysis, with quadratic elements. Approximately 60 000 elements (200 000 nodes) consisting of about: 39 000 hexahedrons; 12 000 pyramids; 7 000 tetrahedrons; 2 000 prisms. Element types used in these analyses can be found as SOLID186 and SOLID187 in Ansys manual [S7]. The results have been also analyzed for comparison with PMD software [S8]. In the disc, zero displacement was applied in the opposite face to the contact with the anvil. Boundary conditions in the anvil were applied with respect to the local coordinate system, where the *z*-axis is oriented along the rotational axis (*c*-axis). In this case, displacement was allowed only in *z*-direction and the force was applied on the top face of anvil along the same direction. Augmented Lagrange formulation [S9] was used for contact, using friction coefficient of 0.2 for sapphire to sapphire, with symmetric behavior (no difference between master and slave faces (contact and target in Ansys).

a)

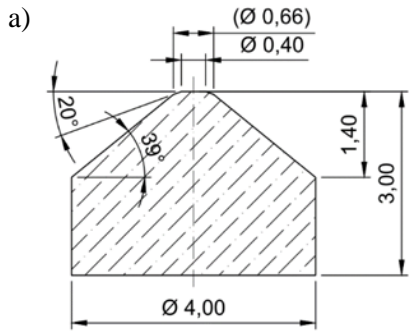

b)

$$
\begin{pmatrix}
c_{11} & c_{12} & c_{13} & c_{14} & 0 & 0 \\
c_{12} & c_{11} & c_{13} & -c_{14} & 0 & 0 \\
c_{13} & c_{13} & c_{33} & 0 & 0 & 0 \\
c_{14} & -c_{14} & 0 & c_{44} & 0 & 0 \\
0 & 0 & 0 & 0 & c_{44} & c_{14} \\
0 & 0 & 0 & 0 & c_{14} & 1/2(c_{11}-c_{12})
\end{pmatrix}
$$

**Figure S9**. a) Geometry and dimensions of the sapphire anvil. b) Elastic matrix with elastic moduli: $c_{11}$ = 4.902 $10^{11}$ Pa, $c_{44}$ = 1.454 $10^{11}$ Pa, $c_{13}$ = 1.130 $10^{11}$ Pa, $c_{33}$ = 4.902 $10^{11}$ Pa, $c_{12}$ = 1.654 $10^{11}$ Pa, $c_{14}$ = -0.232 $10^{11}$ Pa [S10].

**Atomic Force Microscopy (AFM)**

The AFM images were obtained using Dimension Icon microscope (Bruker) operating in Peak Force Tapping mode using ScanAsyst-Air probes (stiffness 0.2-0.8 N/m, frequency ~80 kHz). No treatment apart from line subtraction (retrace) to remove the tilt has been performed. Figure S10 shows AFM images (topography in the left column, adhesion between the tip and the sample in the right column) of three areas: two bilayers (top and middle) and one monolayer (bottom). As can be seen, it is rather straightforward to identify regions with and without graphene in the monolayer through the differences in adhesion between the substrate and the sample. The height difference measured then in the topography channel gives ~0.6 nm. It is, however, much more difficult to pinpoint areas with neighbouring monolayer and bilayer CVD graphene. In certain cases, holes in the top layer are unveiled thanks to visible folds at the side of the hole, as in Figure S10a and c (the boundaries of the holes are marked by blue lines and the respective fold in (a) by a blue arrow). No changes can be observed in the adhesion channel. The average difference in height between the green and red regions gives 0.64 ± 0.25 nm, very similar to the spacing between the substrate and the bottom layer. The large standard deviation is due to the high amount of impurities and folds present in the transferred layers, and the even higher concentration of those close to the hole boundaries.

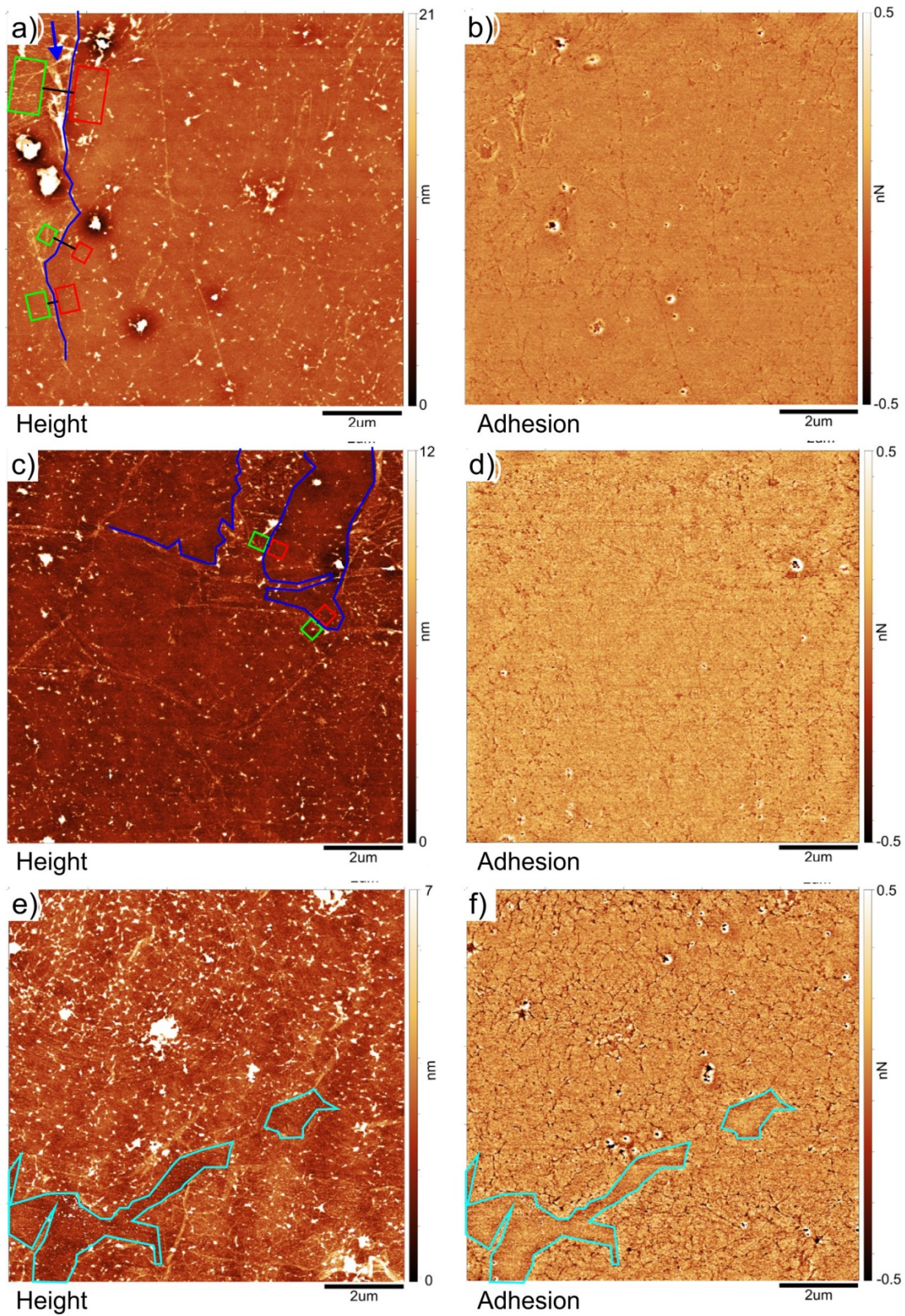

**Figure S10.** The AFM topography(a, c, e) and adhesion (b, d, f) images of bilayer (a-d) and monolayer (e-f) CVD graphene.

**Comparison of G/2D ratio evaluated as amplitudes and integrated areas**

To retain quantitative consistency in the G/2D ratio evaluation throughout the main text, amplitudes of the peaks were used (from fitting with Lorentzian shapes). As shown below, there is only negligible qualitative difference between amplitude and integrated area plots. Figure 3 from the main text is plotted in the top row in Figure S11, and Figure 4 from the main text forms the top row in Figure S12 (i.e. the amplitude ratio A(G)/A(2D)), while the bottom plots in both figures show the ratio for integrated areas (I(G)/I(2D)). The ratio of integrated areas is ~2 times lower consistently for all data points, but otherwise the factor of G band enhancement remains the same.

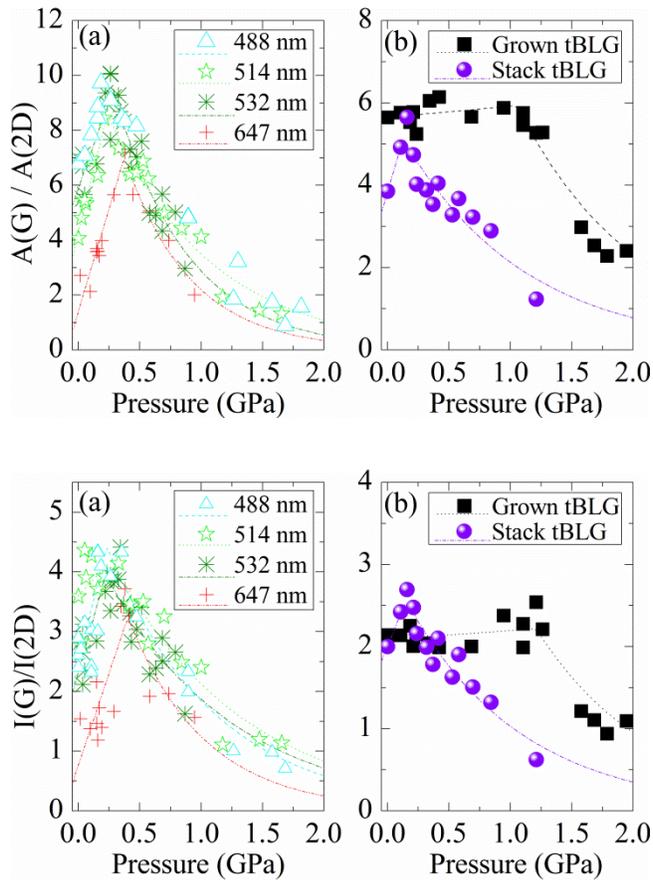

**Figure S11.** Evolution of the G/2D amplitude ratio (top) and integrated area (bottom) with increasing stress: (a) Labeled tBLG with rotation angles of 13.0, 12.3, 11.9 and 9.8° excited with 488.0, 514.5, 532.0 and 647.1 nm laser wavelengths, respectively, to achieve G band enhancement in each sample. (b) $^{12}$C CVD BLG (squares) and $^{12}$C/$^{12}$C stack BLG (circles); both measured in a region resonant with 488.0 nm excitation wavelength. Dotted lines are guides for the eye.

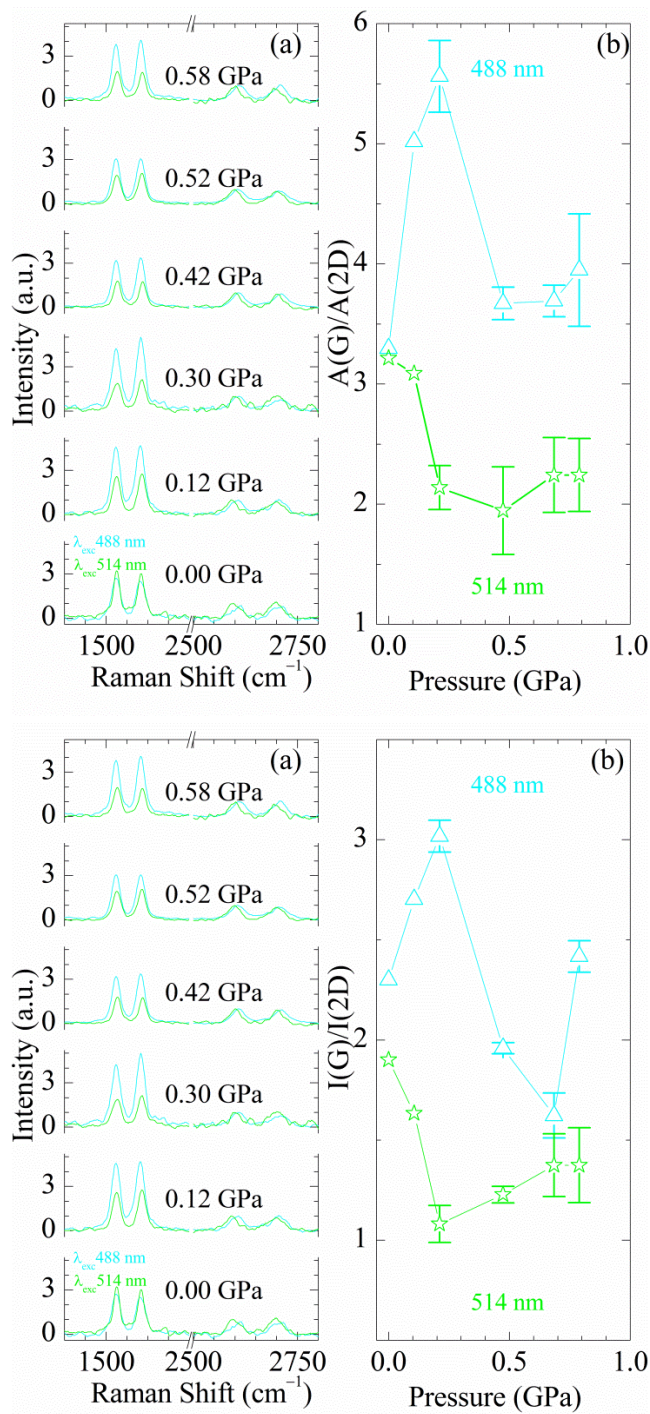

**Figure S12.** Evolution of Raman spectra of $^{13}C/^{12}C$ tBLG with compression up to 0.6 GPa. (a) The same tBLG grain is measured with 488.0 (blue spectra) nm and 514.5 nm excitation wavelengths (green spectra). (b) G/2D ratios as amplitudes (top) and integrated areas (bottom) from (a) as a function of increasing stress.